# Equity and Artificial Intelligence in Education:
# Will "AIEd" *Amplify* or *Alleviate* Inequities in Education?

Kenneth Holstein* and Shayan Doroudi*

## INTRODUCTION

With increasing awareness of the societal risks of algorithmic bias and encroaching automation, issues of fairness, accountability, and transparency in data-driven AI systems have received growing academic attention in multiple high-stakes contexts, including healthcare, loan-granting, and hiring (e.g., Barocas & Selbst, 2016; Holstein, Wortman Vaughan, Daumé III, Dudik, & Wallach, 2019; Veale, Van Kleek, & Binns, 2018). While questions of how to design more transparent and accountable systems have received some attention within the academic field of AI in Education (e.g., Bull & Kay, 2010; 2016; Conati, Porayska-Pomsta, & Mavrikis, 2018; Holstein et al., 2019; Shum, 2018), issues of fairness and equity in educational AI (AIEd) systems have received comparatively little attention (Blikstein, 2018; Ferguson, 2019; Holmes, Bialik, & Fadel, 2019; Holstein & Doroudi, 2019; Shum & Luckin, 2019).

The development of AIEd systems has often been motivated by their potential to promote educational equity and reduce achievement gaps across different groups of learners – for example, by scaling up the benefits of one-on-one human tutoring to a broader audience (e.g., Kizilcec, Saltarelli, Reich, & Cohen, 2017; O'Shea, 1979; Reich & Ito, 2017; VanLehn, 2011; 2016), or by filling gaps in existing educational services (e.g., Madaio et al., 2020; Saxena, Pillai, & Mostow, 2018; Uchidiuno et al., 2018). Given these noble intentions, why might AIEd systems have inequitable impacts?

In this chapter, we ask whether AIEd systems will ultimately serve to *Amplify* **I**nequities in **Ed**ucation, or alternatively, whether they will help to *Alleviate* existing inequities. We discuss four lenses that can be used to examine how and why AIEd systems risk amplifying existing inequities: (1) factors inherent to the overall *socio-technical system design*; (2) the use of *datasets* that reflect historical inequities; (3) factors inherent to the underlying *algorithms* used to drive machine learning and automated decision-making, and (4) factors that emerge through a complex *interplay* between automated and human decision-making. Building from these lenses, we then outline possible paths towards *more equitable futures* for AIEd, while highlighting debates surrounding each proposal. In doing so, we hope to provoke new conversations around the design of equitable AIEd, and to push ongoing conversations in the field forward.

---

* These co-first authors contributed equally.

## PATHWAYS TOWARD INEQUITY IN AIED

We begin by presenting four lenses to understand how AIEd systems might amplify existing inequities or even create new ones (cf. Buckingham Shum, 2018). While each lens provides a different way of examining pathways towards inequity in AIEd, all are pointed at the same underlying socio-technical system. Figure 1 provides a coarse-grained overview of the broader social-technical systems in which AIEd systems are embedded, and some of the components we will refer to in the four lenses. The accumulated, collective decisions of designers, researchers, policy-makers, and other stakeholders shape these systems' designs. In addition to *using* or being *affected by* AIEd systems, on-the-ground stakeholders such as students, teachers, or school administrators may also play a role in shaping their designs; whether *directly*, through participatory design processes, or *indirectly* through the passive generation of training data while interacting with an AIEd interface. In turn, decisions regarding what data is used to shape an AIEd system's design (e.g., when used as training data for use with machine learning methods) can shape an AIEd system's algorithmic behavior (e.g., instructional policies learned from data). This relationship is often bidirectional: the AIED system's algorithmic behavior also determines and constrains the kinds of data that are subsequently generated via user interactions.

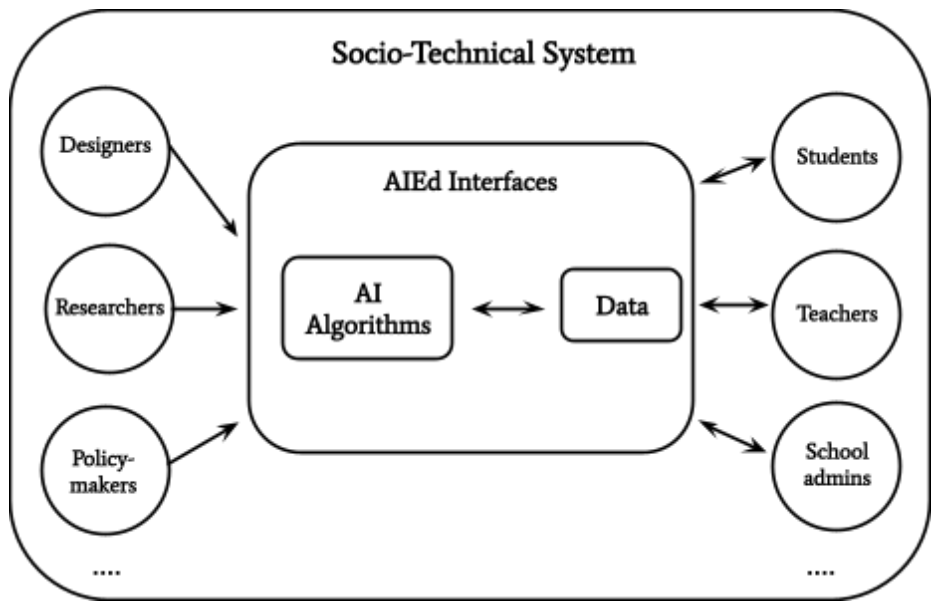

**Figure 1.** High-level overview of AIEd systems and the broader socio-technical systems in which they are embedded.

**Designing more equitable AIEd technologies requires reasoning at multiple levels of abstraction, rather than only thinking about one level in isolation**. However, looking at these challenges through different lenses at times can reveal different kinds of potential solutions. Thus, these four lenses differ primarily in *where they place their focus*, with each dedicating

higher resolution to particular components of the overall socio-technical system. Indeed, different research communities have tended to tackle issues of fairness and equity in AI systems, both within and outside the area of education, through different sets of lenses. While AI and machine learning communities have historically studied these issues primarily through the second (Data) and third (Algorithmic) lenses, fields such as human–computer interaction (HCI) and the learning sciences have largely approached these issues through the first (System) and fourth (Human–Algorithm) lenses. As we discuss below, designing more equitable AIEd systems will likely require meaningful integration across all four lenses.

Kizilcec and Lee (this volume) discuss the particular concern of algorithmic fairness in educational systems in greater depth, including an exposition of technical definitions of fairness, which we do not discuss in this chapter. Baker and Hawn (2021) also provide an overview of the sources of algorithmic bias throughout the machine learning pipeline, providing examples of pathways that go beyond the ones discussed here. We see issues of algorithmic fairness as constituting an important subset of the pathways that could lead to inequity in AIEd systems that we discuss in this chapter. Kizilcec and Lee (this volume) characterize three steps in the design and use of an algorithmic system that can result in unfairness: measurement, model learning, and action. These roughly correspond to our data, algorithmic, and human–algorithm lenses respectively. However, since we take a broader view of inequity in AIEd systems, each of these three lenses also include ways in which inequity can emerge even in the absence of using data-driven techniques. Overall, these recent overviews of algorithmic fairness in education are complementary to the broader discussion of equity in AIEd presented in this paper.

**Lens 1 (System): Factors Inherent to the Overall Socio-Technical System Design**

We first discuss factors related to the design of AIEd technologies *other than* their use of AI algorithms that may contribute to inequities in education. These factors are highlighted in Figure 2. They include the components of a technology that does not explicitly involve the use of AI, such as the user interface, the learning theories and pedagogical principles underpinning the system, and the way domain knowledge is represented in the system. This lens also considers all the relevant factors of the broader socio-technical system in which the technology lives, which includes various human stakeholders, values, beliefs, business models, the context in which the system is used, and so on.

Under this lens, one major source of inequity of AIEd technologies lies in *disparities of access*, where a technology is more accessible to certain groups of learners than others. For example, costlier technologies may be more likely to be adopted by wealthier schools or families. Although the cost of computers (including mobile devices such as phones) has greatly declined, computer access remains a source of disparity. A 2015 survey found that 5% of families in the US with children under the age of 18 do not have *any computer* in their household (desktop, laptop, or handheld device) and 15% have no internet subscription (Ryan and Lewis, 2017). Moreover, these numbers are lower for underrepresented minorities than for White and Asian families. Rideout and Katz (2016) found that even though the percentage of families without any kind of Internet access is relatively small, many low- and moderate-income families with nominal Internet access were still found to be under-connected in one or more ways (e.g., slow Internet access, computer sharing, mobile-only access, and limited data plans). Under-connectivity can in turn affect the degree to which children use digital technologies for educational purposes; for example, Rideout and Katz (2016) found that only 35% of children with mobile-only access would often use the Internet for interest-driven learning as compared to 52% of those who had desktop or laptop access.

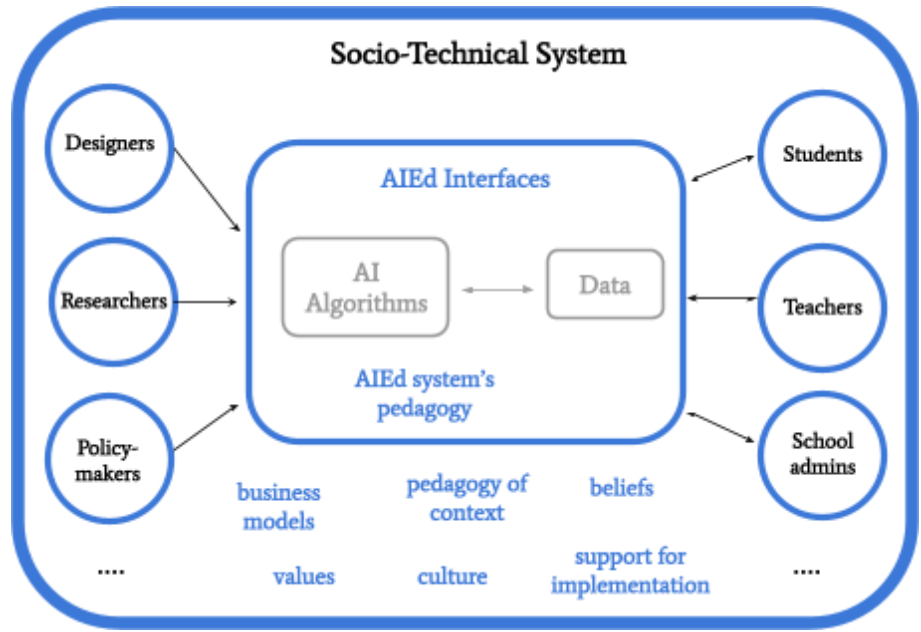

**Figure 2.** Lens 1 (System) focuses on the overall socio-technical system design, rather than the actual AI or data-driven aspects of an AIEd system. In this figure and in subsequent figures, the components highlighted in blue show the focus of the current lens, while those greyed out are out-of-focus.

Yet even if all students have physical access to a given technology, disparities of access may remain. For example, an AIEd system that provides guidance (e.g., on-demand hints or scaffolding questions) only in English will limit access to non-English speakers, and may serve to disadvantage non-native versus native English speakers. Where AIEd systems assume use of a mainstream dialect, students from minoritized groups who use other dialects at home may be disadvantaged. For example, Finkelstein et al. (2013) showed that African American students displayed more scientific reasoning when taught by a virtual avatar that spoke entirely in African American Vernacular English (AAVE), compared with an avatar that spoke Mainstream American English (MAE). Such disparities can also arise when educational *content* is implicitly tailored toward particular cultural contexts. Chipman et al. (1991) found that the familiarity of situations described in math word problems affected performance on those problems, even when the underlying mathematical structure was held constant. Moreover, unless AIEd technologies are explicitly designed with accessibility in mind, these technologies risk accelerating learning for some groups of learners, while decelerating learning for others; this can happen, for example, if a school implements the technology to replace a previously accessible activity with one that is less accessible (Guo, Kamar, Wortman Vaughan, Wallach, & Morris, 2019; Rose, 2000; Wen, Amog, Azenkot, & Garnett, 2019).

Inequities due to the overall socio-technical system design extend well beyond disparities of access. Reich and Ito (2017) synthesize a body of research showing that even when schools and individual learners have *equal access*, "new technologies tend to be used and accessed in unequal ways, and they may even exacerbate inequality" (p. 3). Different schools and teachers will use the same technologies in different ways, for example due to differences in values and beliefs surrounding education and technology. For example, Rafalow (2020) found that teachers in schools with different demographic compositions adopted different attitudes toward students' digital literacy skills and expressions based on (racial) stereotypes about the student body. Notably, schools that are better-resourced and that serve students from more privileged backgrounds tend to use technologies in more innovative ways. While this is not solely a technical challenge, designers of AIEd technologies should be aware of systematic differences in the ways these technologies are implemented or adopted, and consider ways of measuring, responding to, and designing for such differences. Even when students in the same school have similar levels of access *and* comparable levels of digital literacy, Sims (2014) observed that different social groups might form "differentiating practices" around the use of technology, which could exacerbate social divisions. Further, Reich and Ito (2017) identified that providing open access to educational resources (e.g., massive open online courses), *does not* mean everyone will benefit from those technologies equally. MOOCs are disproportionately used by students with high socio-economic status, many of whom already have advanced degrees. This could be due to socio-cultural barriers that prevent individuals from using certain technologies, even when they are free. However, Kizilcec et al. (2017) showed that simple social belonging and value-affirmation interventions could increase the persistence and completion rate of MOOCs for students from less developed countries, potentially eliminating the gap in MOOC completion between students in less versus more developed countries. Similarly, designers of AIEd technologies should consider interventions to reduce gaps in who *chooses to use* their technology. Simply improving access is not enough.

A key factor that may contribute to inequities in AI-supported education is the "social distance between developers and those they seek to serve" (Reich and Ito, 2017). Even in cases where teams explicitly design technologies to help underserved populations, if the design process is not guided by representative voices from those populations, the resulting technologies may fail to serve the needs of those populations or may even amplify existing equity gaps (as discussed further in the next section).

So far, we have largely discussed factors that apply to educational technologies in general, not only AIEd systems. However, common design features of AIEd systems have been identified which may contribute to inequities in use. For instance, many AIEd systems, such as intelligent tutoring systems, are designed for *one-on-one interaction* with a student. One-on-one interaction is assumed not only in these systems' interface designs, but also in the choices of the AI algorithms that these systems rely upon to adapt and personalize instruction (e.g., Bayesian Knowledge Tracing; Corbett & Anderson, 1995). If the system is used in a different way in practice, it may make decisions that are suboptimal or even harmful to students (see our third lens below, as well as Olsen, Aleven, & Rummel, 2015). Yet research has shown that in certain cultural contexts, students tend to work with these technologies *collaboratively* rather than individually (Ogan, Walker et al., 2015; Ogan, Yarzebinski, Fernández, & Casas, 2015).

Another factor is the lack of flexibility or *adaptability* of AIEd systems. Since AIEd technologies are often complex systems with many interlocking components, these systems tend to permit limited end-user customization. In particular, AIEd systems rarely support flexible customization options, to help teachers adapt these systems to the needs of their local contexts (Nye, 2014). On the other hand, simpler educational technologies that do not rely (extensively) on AI may lend themselves to such customization more easily, by default. For example, in lieu of advanced forms of adaptivity, the ASSISTments system is designed to support *teacher-driven* adaptability, allowing teachers to create or customize their own assignments so they can better meet the needs of their students (Heffernan & Heffernan, 2014). This support for teacher-driven adaptability means that teachers are not required to rely on pre-set assignments with which they may be unfamiliar, or which may be poorly fit for their contexts (e.g., by presenting word problems with which their students cannot connect).

**Lens 2 (Data): Use of Data that Reflect Historical Inequities**

AIEd systems have great potential to address inequities in education, for example by filling gaps in the educational services otherwise available to learners (Madaio et al., 2020; Saxena et al., 2018; Uchidiuno et al., 2018). However, insofar as these technologies are shaped or driven by historical data, they risk perpetuating or even amplifying any social inequities reflected in these data (Mayfield et al., 2019; Selbst et al., 2019; Veale et al., 2018). This third lens considers the ways in which biases and inequities in the ways in which data was collected or generated can propagate in AIEd systems. The focus of this lens is depicted in Figure 3.

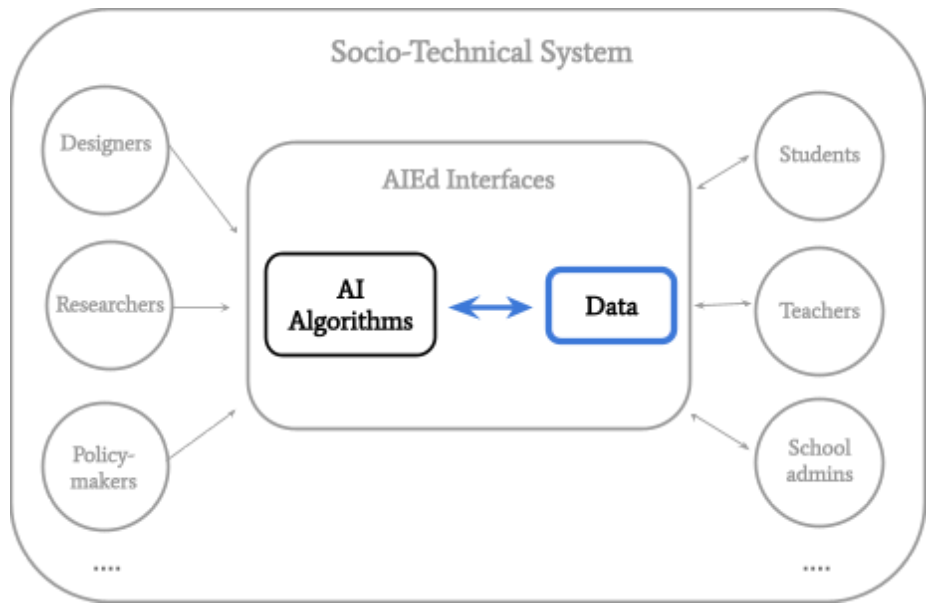

**Figure 3.** Lens 2 (Data) highlights decisions regarding the data that an AIEd system relies upon (including the ways in which this data is used or generated). Designing datasets or data-generating mechanisms that support more equitable outcomes is a central goal.

The risks of "scaling up" inequities via data-driven approaches have inspired new research communities focusing on fairness in machine learning (e.g., the ACM Conference on Fairness, Accountability, and Transparency), which have begun to influence research on educational data mining and learning analytics. Through this second lens, inequities may arise through a number of possible pathways.

The first pathway highlights the fact that biases in data need not arise only when machine learning or data-driven techniques are explicitly used. Here, we consider the impact of social distance discussed under the first lens through this data-centric lens, centering the roles of *sampling processes and biases* in the design of a system. Even if a given AIEd system's behavior is not shaped through machine learning, its design process may still be guided by the use of data. This data may often consist of observations from a limited set of contexts and feedback from a limited set of stakeholders. Meaney & Fikes (2019) discuss "Early-adopter Iteration Bias" in the design of learning analytics systems and other educational technologies—the notion that the educational practitioners and students most likely to participate at the earliest stages of the design process tend to represent a relatively privileged subpopulation. In the presence of such recruitment biases, the designs of novel educational technologies may be skewed towards the needs of these early adopters, while failing to reflect the needs of marginalized, underserved, and otherwise risk-averse populations (who may participate more heavily in later design cycles).

As a second major pathway, when an AIEd system's behavior is developed in a data-driven manner, inequities may arise due to *misalignments* between system developers' educational goals, the datasets used to train and evaluate models, and the dynamics of the contexts in which these systems will eventually be used. If particular learner subpopulations or educational contexts are overrepresented in datasets used to train and evaluate models offline, the resulting models may encode associations that fail to generalize to other learners and contexts, and which may even cause harm if deployed in these contexts (Ocumpaugh, Baker, Gowda, Heffernan, & Heffernan, 2014). For example, the use of proctoring software during the COVID-19 pandemic has raised several equity concerns, including difficulties in identifying the faces of students of color, presumably due to the biases in the datasets that the facial recognition technologies were trained on (García-Bullé, 2021; Teninbaum, 2021).

Moreover, in cases where AIEd systems' assessment or pedagogical models are learned from the ratings or observed behaviors of human decision-makers (e.g., school administrators, teachers, parents, or students), the resulting models may serve to scale up not only beneficial practices, but also *undesirable biases* exhibited by these educational stakeholders (as discussed further under our fourth lens below). For example, it is known that K-12 teachers often exhibit biases in assessing the written work of students from marginalized populations (representing mismatches between teachers' educational goals versus their actual assessment practices). Furthermore, prior work has demonstrated that when human teachers' scores are uncritically used as the "ground truth" for student performance, such assessment biases can easily be encoded into automated essay scoring systems used at scale (e.g., Madnani, Loukina, von Davier, Burstein, & Cahill, 2018; Mayfield et al., 2019). As discussed in the next lens, the extent to which a given AIEd system serves to propagate or even amplify historical inequities may depend in part on the choices of *models and algorithms* used to develop these systems.

A third major pathway is discussed below under our fourth lens (Human–Algorithm). When machine learning based AIEd systems are deployed in real-world educational contexts, historical biases may not only replicate but also *compound* via feedback loops.

**Lens 3 (Algorithmic): Factors Inherent to the Underlying Algorithms**

AIEd algorithms are often implicitly designed to reduce equity gaps, for example, by attempting to personalize instruction for each learner. Giving students the right amount and types of instruction at the right time can mitigate concerns around one-size-fits-all instruction that tend to amplify existing achievement gaps. Nonetheless, due to various factors, the use of AI algorithms may fall short of this vision by engendering inequitable outcomes for learners through a number of possible pathways. This lens focuses on the use of AI algorithms—including the ways in which these algorithms utilize or are shaped by data—as depicted in Figure 4.

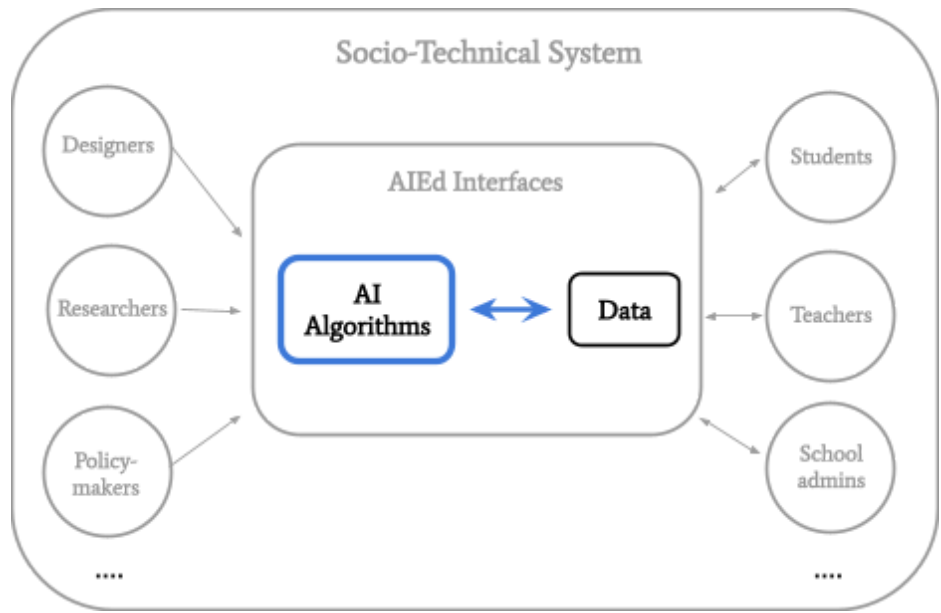

**Figure 4.** Lens 4 (Algorithmic) highlights design decisions regarding an AIEd system's underlying AI algorithms (including the ways in which these algorithms use or are shaped by data). Designing algorithms that support more equitable outcomes is a central goal.

First, as discussed above, machine learning algorithms will tend towards further propagating, or even exacerbating, existing biases present in historical datasets unless explicitly designed to counteract these biases. For example, when fitting a model to an automated essay scoring dataset where teachers' essay grades systematically encode certain biases, the model will naturally make predictions that are biased in similar ways. Note that such scenarios can play out *even if no demographic variables are used* in a machine learning model, because other variables can (and often will) correlate with demographic variables in non-obvious ways (Barocas & Selbst, 2016; Datta et al., 2017, Dwork et al., 2012; Kilbertus et al., 2017). For example, Gardner, Brooks, & Baker (2019) found that across a variety of MOOC datasets and different machine learning models used to predict dropout rates, the models exhibited disparities in accuracy across males and females, even though none of the models used gender or other demographic variables as explicit features. Although the source of such disparities may be understood under our second lens to lie in the choices of data used to train these models, particular choices of models and learning algorithms may be more prone to amplify such biases than others (Elzayn et al., 2019).

Even if theoretically *there were no undesirable historical biases* in a dataset, machine learning algorithms could still be unfair when the amount of data we have for different demographic groups vary. In particular, most machine learning algorithms will be *inherently biased* against minority groups in our data (Hardt, 2014). This is because the algorithm will try to optimize overall accuracy of a model, and it can achieve that accuracy by tailoring its predictions to that of a majority group (at the expense of minority groups), it will do so. For example, a learning analytics algorithm that detects "at-risk" students trained on data from a majority-white college can be very accurate overall, by learning how to make nuanced predictions for the white students and possibly making less nuanced predictions for Black students or Native American students for whom it may have much less data. To reiterate, this can result simply from the fact that minority groups will naturally be less represented in a dataset where they are in a minority, rather than systematic historical biases (although in practice, certain groups may be undersampled in available datasets due to historical biases). In some cases, this concern might be resolved by intentionally collecting more data from minority groups, even if that is not representative of the population.

Machine learning algorithms can also contribute to inequitable decision-making due to *model misspecification*. This is especially relevant when fitting statistical models of student learning, since learning is a complex, unobservable phenomenon, and any simple model is bound to be an imperfect representation of the true underlying processes. For example, Doroudi and Brunskill (2019) examined contexts where a student model is used for mastery learning (i.e., to decide how much practice a student should receive on each of a set of targeted skills or knowledge components). The authors demonstrated two major kinds of model misspecification that can cause inequitable outcomes. First, if learners learn at different rates (i.e., there are faster and slower learners), but the student model is not individualized (i.e., a single model is fit to aggregate data from faster and slower learners), then slower learners may be more likely to receive less practice than they need as compared to fast learners. Second, if the functional form of the model is fundamentally misspecified, then slower learners could again face disadvantages (receiving suboptimal practice). In particular, the authors showed that if in reality, students learn according to the Additive Factor Model, but a Bayesian Knowledge Tracing model is used for mastery learning (as is typically the case in intelligent tutoring systems), then *even if* separate models are fit for slower versus faster learners, inequitable outcomes across these groups persist as a result of the underlying model misspecification.[1] Therefore, in order to ensure equitable outcomes, AIEd system developers should not only strive to develop student models that more faithfully capture student learning processes, but also aim to develop algorithmic interventions that are robust to potential model misspecification.

Finally, AIEd algorithms can also be inequitable *even when they are not data-driven*. For example, using a heuristic for mastery learning, such as assuming students have learned a skill when they display *N* consecutive correct responses for that skill (Kelly, Wang, Thompson, and Heffernan, 2016; Hu, 2011), could also contribute to inequitable outcomes (Doroudi & Brunskill, 2019). Implicitly, such a heuristic is making assumptions about student learning that may not actually hold in practice in all contexts (Doroudi, 2020). In general, algorithms that make different decisions for students based on their performance, while key to providing adaptive instruction, can be susceptible to such risks. For example, a recently proposed game-theoretic mechanism for incentivizing students to accurately self-assess their work (Labutov & Studer, 2016) can *inherently* lead to artificial deflation of lower-performing students' grades. That is, lower-performing students could receive much lower grades than they would otherwise receive, if they were not asked to self-assess their work. Outside of the research setting, similar scoring rules are being used to grade students in several university courses on decision analysis (Bickel, 2010). Yet researchers have not discussed the consequences of these incentive mechanisms with respect to equity in educational outcomes.

---
[1] Misspecifying the functional form of a student model may not always result in worse results for low-performing students, so this should be studied on a case-by-case basis.

**Lens 4 (Human–Algorithm): Interplay between Automated and Human Decision-making**

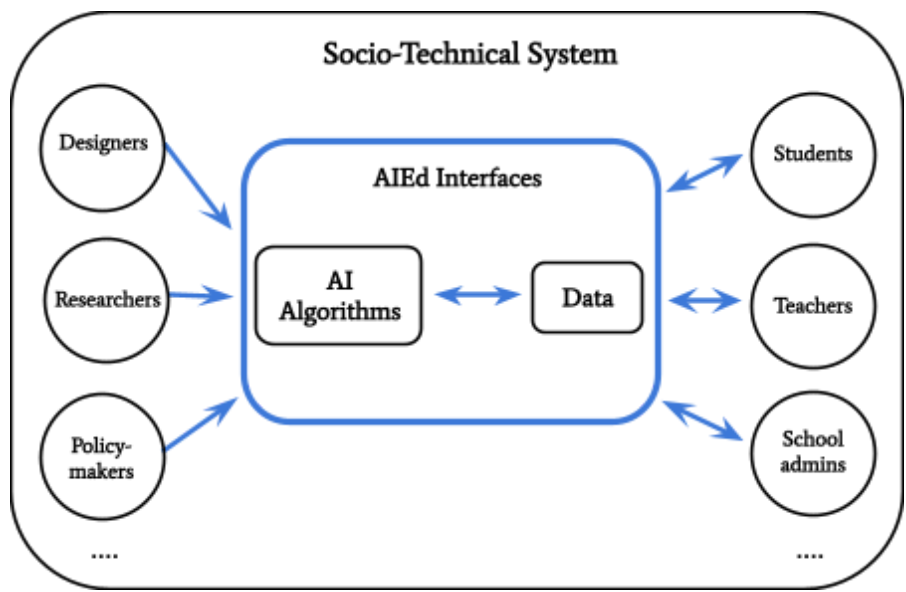

**Figure 5.** Lens 4 (Human-Algorithm) highlights the complex *interactions* between AIEd systems and human decision-makers. Under this lens, designing to shape these (often bidirectional) interactions in ways that support more equitable outcomes is a central goal.

When AIEd systems are used in real-world educational contexts, they do not act upon students in a vacuum. Rather, their impacts result from a complex interplay between the AI systems themselves, the human teams who design and develop them, and the human decision makers who use these systems (e.g., teachers, school administrators, and students; Holstein et al., 2018b; Holstein, Wortman Vaughan, et al., 2019; Mayfield et al., 2019; Madaio, Stark, Wortman

Vaughan, & Wallach, 2020). As such, achieving more equitable futures for AI in Education requires carefully designing for this interplay (cf. De-Arteaga, Fogliato, & Chouldechova, 2020; Holstein et al., 2019)—moving beyond a focus on AIEd interfaces, datasets, or algorithms in isolation—as depicted in Figure 5.

In cases where an educational AI or learning analytics system does not necessarily encode harmful *algorithmic* biases (e.g., where the system does not present disparities in predictive accuracy across subgroups of students), it is still possible that real-world *use* of the system may systematically contribute to inequitable outcomes. For instance, consider two alternative designs of a decision-support tool for classroom teachers that both rely upon the same underlying student modeling algorithms, yet present the outputs of these algorithms to teachers in different ways. Despite the use of common algorithms, different methods of presentation may interact with teachers' existing decision-making biases in very different directions (see Holstein et al., 2018a; 2018b; Martinez-Maldonado, 2013; van Leeuwen 2015; van Leeuwen & Rummel, 2020). If a tool design leads teachers to *challenge* their existing beliefs about individual students' abilities and potential for growth, then this design has the potential to nudge teachers towards more equitable practices (Holstein et al., 2018b). By contrast, if a tool design has the primary effect of making teachers feel *validated* in their existing beliefs, even when these beliefs are incorrect, then this design risks maintaining or even amplifying inequitable patterns of teacher behavior (Holstein et al., 2018a; 2019; van Leeuwen, 2015).

Such interactions between human and machine biases may have compounding effects – for better or worse – in the presence of human–AI *feedback loops* in learning or decision-making (cf. Lum and Isaac, 2016; Elzayn et al., 2019; Veale et al., 2018). For example, in the context of predictive policing, Lum and Isaac (2016) showed that PredPol, a popular predictive policing software, reinforced existing biases in the dataset and targeted Black people at disproportionately high rates (twice as high as white people, even though drug use was roughly equal between the two groups); this is a typical inequitable outcome as per the second lens. However, they further showed via simulation that if police are more likely to find crime in areas where they are actively policing (as a result of PredPol's predictions), which in turn affects the crime rates in the data used by the algorithm, then the algorithm will become further biased in a feedback loop that amplify existing biases.

In cases where deployed AI systems both *shape* the behaviors of learners and educators (e.g., by presenting particular predictions, recommendations, or behavioral nudges) and are *shaped by* these evolving behaviors (continuously learning and adjusting based on incoming educational data), the dynamics of the resulting educational system can be challenging to anticipate. Given a longstanding focus on the development of data-driven, self-improving systems for education (Doroudi, Aleven, & Brunskill, 2019; O'Shea, 1979), measuring, modeling, and designing for such cyclical interactions should be a central focus of future work. If intentionally designed as such, these cycles may have the potential to serve as positive, regulative loops (cf. VanLehn, 2016)—helping humans to mitigate undesirable impacts of algorithmic decision-making and helping machines to mitigate undesirable impacts of human decision-making.

## TOWARDS MORE EQUITABLE FUTURES FOR AI-SUPPORTED EDUCATION

In the following, we present brief examples of some possible paths towards more equitable futures for AI in Education. We first discuss promising paths towards *avoiding amplification* of existing inequities through AIEd, addressing some of the risks discussed in the last section. After seeing how we might mitigate such negative outcomes, we then return to the promise of AIEd as a means of combating inequity, presenting some possible paths for AIEd to *alleviate* existing inequities in education.

For each path presented, we also present one or more critiques of the proposed path, followed by potential resolutions. In presenting each path in the style of a dialectic (inspired by repeated conversations we have had with AIEd researchers and practitioners), we intend to highlight the challenges that we may face ahead, while also acknowledging their complexity. Our goal is to both provoke new conversations and to push ongoing conversations forward. However, note that the following is *far from an exhaustive set* of promising paths forward or an exhaustive discussion of associated challenges and critiques.

### Pathways to *Avoid Amplifying* Inequities

**Path 1: Invest in developing tools and processes to support AIEd practitioners in developing and deploying more equitable technologies.**
We could further invest, as a field, in research and practical methods for assessing and addressing inequitable impacts of AIEd technologies. For example, we could develop specialized toolkits and organizational processes to support equity-focused design, testing, and deployment in the area of AI-supported education (Cramer et al., 2019; Holstein, Wortman Vaughan, et al., 2019; Madaio, Stark, et al., 2020). AIEd development teams could then *continuously* monitor for unwanted disparate impacts of their technologies across different educational contexts and groups, and iteratively refine these technologies to address such disparities.

> **Critique 1: How will AIEd companies be held accountable?**
> Will companies simply be able to announce that they are auditing for inequitable impacts of their technologies, to improve public relations, but without actually reporting all issues they uncover or addressing these issues? (cf. Bietti, 2020).
>
> **Possible resolution**: We can and should design mechanisms to promote increased accountability and transparency. For instance, support for company-internal auditing should exist side-by-side (and ideally in close coordination with) mechanisms for auditing and regulating AIEd systems from the outside (cf. Cramer et al., 2019; Madaio, Stark, et al., 2020; Raji & Buolamwini, 2019).

**Critique 2**: **In addressing one form of inequity, AIEd developers may inadvertently create others.**
By trying to satisfy a particular notion of fairness (e.g., according to a particular quantitative definition proposed in prior literature), one might actually cause harm in another dimension without realizing it (Holstein, Wortman Vaughan, et al., 2019; Kleinberg, Mullainathan, Raghavan, 2016; Zhu, Yu, Halfaker, & Terveen, 2018).

**Possible resolution**: Nonetheless, we need to try our best. Ultimately, the notions of fairness we choose for a particular application depend on what we, and those we include in the design process, find most salient and important. Where fundamental design trade-offs exist, we *need to take a stance* and make these trade-offs. This means accepting responsibility, which is better than the alternative of denying responsibility. Moreover, by continuously auditing for fairness, we avoid making a small one-off fix that leads to unanticipated long-term side effects.

**Path 2: Design AIEd systems that can clearly communicate their own capabilities and limitations, and hand off control to humans as needed.**
Future AIEd systems can be designed to communicate their uncertainties and potential biases to the humans-in-the-loop (e.g., when reporting analytics), allowing relevant human decision-makers (e.g., teachers and students) to override their recommendations when needed (Amershi et al., 2019; De Arteaga, Fogliato, & Chouldechova, 2020; Fancsali, Yudelson, Berman, & Ritter, 2018; Holstein et al., 2018b; 2019).

**Critique**: **"Lifting the curtain" on AIEd systems will reduce trust.**
By increasing transparency and making people more aware of the boundaries of AIEd systems' capabilities, we risk promoting *under-trust* in these systems. Ultimately, these kinds of efforts will make people less likely to use systems that would otherwise be beneficial to students.

**Possible resolution**: If such systems are designed carefully, with an awareness and understanding of the ways human and algorithmic decision-making interact (e.g., algorithmic aversion or over-reliance), we can support teachers and students in achieving *the right levels of trust* in particular instances. Ideally, AIEd systems should support teachers and students in trusting their predictions and recommendations only when they are *trustworthy*, and should otherwise support teachers and students in second-guessing or overriding such information (cf. Bull & Kay, 2016; Conati et al., 2018; Holstein et al., 2019). Prior research suggests that increased transparency in AI systems can have the overall effect of improving trust rather than diminishing it, at least when paired with corresponding, meaningful options for *user control* (Lee & Baykal, 2017).

**Path 3: Incorporate equity-related outcomes, not just individual student learning outcomes, into the objective functions of AIEd systems.**
"Rich get richer" effects are often observed in classrooms using AIEd systems, where systems behave in ways that systematically benefit students coming in with higher prior domain knowledge *more* than they benefit students coming in with lower prior knowledge (e.g., Doroudi & Brunskill, 2019; Hansen & Reich, 2015; Holstein et al., 2018b; Rau, 2015). As a result, although AIEd systems may show favorable results at the individual student level, at the population-level they risk *further widening achievement gaps*. To address this, we should work towards designing algorithms for AIEd systems that optimize not only for individual student learning outcomes, but should also take into account outcomes at higher levels (e.g., at the class-, school-, district-, or region-level). See Kizilcec and Lee (this volume) for a related discussion of this issue.

> **Critique: We should not suppress advanced students' progress simply to avoid widening achievement gaps across students.**
> A positive framing of the "rich get richer" argument is the notion that AIEd systems can help advanced students reach their full potential faster than they would otherwise. It may not be ethical or desirable to suppress these students' progress in the name of equity. Education is complex, and narrowing achievement gaps is *not our only goal*.
>
> **Possible resolution**: Ultimately, it may be that we may need to make trade-offs. However, we should make sure that we do so consciously and deliberately, rather than making these kinds of trade-offs "by default." As a field, we should critically reflect on our goals for learning, both at the individual and population levels, as well as the tradeoffs we are willing to make among these goals. It might also be the case that we can overcome certain trade-offs through thoughtful design. For example, many AIEd systems that involve adaptive problem selection will allow advanced students to progress to later material in a curriculum as quickly as they can reach this material (such that in the same class, one student might be working on "Chapter 3" while another is already working on "Chapter 25"). However, an alternative design for such technologies, which might have some desirable properties with respect to equity, might instead keep different students relatively synchronized in the curriculum, while providing challenging enrichment opportunities for those students who move through the material more quickly (e.g., allowing advanced students to *deepen* their knowledge of a particular area, but without necessarily supporting them in moving far ahead of others in the curriculum, possibly by having them help tutor their struggling peers).

**Pathways to *Alleviate* Existing Educational Inequities**

**Path 1: Broaden (meaningful) participation in AIEd system design to more equitably serve diverse populations.**
The design of AIEd systems should take into account voices from diverse perspectives that are representative of the students and teachers these systems aim to serve (e.g., in socio-economic background, culture, and language). This could include involving diverse educational stakeholders (e.g., student, teachers, and parents from a range of backgrounds) in a co-design process, or hiring a more diverse team to lead the design of the technology. However, this should not mean teams should hire "token" members of particular demographic groups simply to cross off a checklist (Arnstein, 1969). A constant effort needs to be made to seek out relevant voices that are not being represented, and to ensure that these voices can actually have a *meaningful* impact on AIEd system designs (Holstein et al., 2019; Madaio, Stark, et al., 2020; Young, Magassa, & Friedman, 2019).

> **Critique 1: Students and teachers are not scientific experts.**
> By enabling broad participation, we allow students and teachers with poor knowledge of good educational practices to potentially lower the quality of these systems. For example, what if a teacher wants to design a system that adapts to student "learning styles," an intervention that scientific research suggests may be ineffective or even *harmful* to students (Pashler, McDaniel, Rohrer, & Bjork, 2008)?
>
> **Possible resolution**: By enabling broader participation in the design of these systems, *we do not lose our own agency* as learning scientists, designers, researchers, and developers. Rather, a participatory, collaborative approach acknowledges that there may be much that we do not know, but which stakeholders know quite well (e.g., regarding their day to day lived experiences), and that we similarly hold considerable knowledge that other stakeholders may lack. Our challenge is then to figure out how best to balance between various stakeholders' desires and perceived needs, and existing scientific or design knowledge. From this perspective, we are empowered by knowledge of different stakeholder groups' desires, boundaries, and challenges, not limited by this knowledge. Understanding students' and teachers' perspectives and experiences can also help in anticipating whether and how an AIEd technology might be implemented across different educational contexts. For example, if teachers believe firmly in learning styles, they may adopt a certain technology in a way that attempts to tailor instruction to learning styles. It is advantageous for AIEd developers to know about both positive and negative adoption patterns, and incorporate that knowledge in the design of such systems.

**Critique 2**: **Can broadening participation yield discriminatory outcomes?**
In some cases, truly taking into account the needs of different groups of students, teachers, and educational contexts may require designing different technologies for different groups of users. Even if the technologies are optimized to serve the needs of these different groups, such *disparate treatment* might ultimately serve to stigmatize members of these groups.

**Possible Resolution**: Disparate treatment is already ubiquitous in education, and is often necessary in avoiding *disparate outcomes* across groups. For example, mastery learning, and in fact, *all adaptive systems,* enact forms of disparate treatment. If a student is struggling, we often wish to give them more opportunity to succeed than a student who quickly masters the same material. Ultimately, we need to design interventions with members of these communities to ensure that these targeted interventions are viewed as helpful, but not stigmatizing or discriminatory.

**Path 2: Fill existing gaps in educational services with AIEd technologies**
AIEd systems can fill gaps, providing educational services where few or none previously existed; for example, by offering instruction in the midst of a multi-week teachers' strike, as in the Madaio et al. (2020) case study.

**Critique: AIEd will function as such an attractive band-aid that we will neglect to heal the underlying wounds.**
The presence of such gap-filling AIEd systems may be used as an "excuse" by policymakers not to invest in higher-quality, longer-term fixes for social problems surrounding education (e.g., the presence of these technologies might ultimately lead to funding cuts for teachers and public schools or to increasing class sizes). Similarly, already under-resourced schools might be more likely to adopt this option to cut costs, potentially exacerbating existing inequities (Watters, 2014).

**Possible resolution**:
This is a serious critique that should not be taken lightly. AIEd systems can do enormous good by filling gaps, but ultimately AIEd researchers, practitioners, and policy-makers share a responsibility to ensure they do not have a net negative impact on society. Our community should continue to discuss, with heightened urgency, what we can do to ensure our work is not used as "band-aids" or excuses not to address deeper problems in our education systems (Blikstein, 2018).

**Path 3: AIEd as a force to overcome harmful human biases.**
As discussed above, AIEd systems have many biases, but many of the paths toward inequity are actually rooted in human biases, including biases that teachers might have (e.g., teachers' beliefs influencing how they use educational technology in lenses 1 and 4 and teachers' biases encoded in data used by AIEd technologies, such as automated essay graders, in lens 2). AIEd systems can actually serve as a tool to push against human educators, as needed, to help them notice and overcome their own biases.

> **Critique: If AI systems can encode harmful biases, how can they be used to help humans overcome harmful biases?**
> By using AIEd systems to nudge teachers towards more equitable decision-making, we may inadvertently do the opposite. For example, AIEd systems may incorrectly push against teacher biases that actually represent helpful heuristics (e.g., based on rich contextual knowledge to which the teacher has access, but the AI does not), or may even push their own harmful biases onto teachers.
>
> **Possible resolution:** AIEd systems might be designed with an awareness of their own fallibility, and avoid offering confident prescriptions for teacher action, as described in Path 1 under "Pathways to Avoid Amplifying Existing Inequities." Even though an AIEd system may encode some harmful biases of its own, these biases may differ from those held by a human educator. By calling teachers' attention to *discrepancies* between their own judgments and those of an AIEd system, the system may promote productive teacher reflection and learning (cf. An et al., 2019; An, Holstein, et al., 2020; De-Arteaga et al., 2020; Holstein et al., 2019). AIEd systems (or system developers) can in turn take feedback from teachers to help pinpoint and mitigate biases that the AIEd system may have.

## CONCLUSIONS

The use of artificial intelligence in education represents one possible means towards improving educational systems and reducing existing inequities. However, as with any attempt to address the wicked problems of education, AIEd systems can easily have unintended impacts, amplifying existing inequities rather than alleviating them. We have described multiple lenses through which to understand how this can occur, each of which focuses on a different aspect of the complex socio-technical system within which AIEd technologies operate. Ultimately, only by examining the many ways in which AIEd systems can amplify inequities can we realize the true potential of AIEd to *alleviate* inequities in education. Looking ahead, we must foster honest, critical discussions around the many challenges that lie in our path. It is our hope that this chapter serves to further ongoing conversations and spark new ones within the area of AI-supported education.